\documentclass[
    ,final            % use final for the camera ready runs
  ]
  {aipproc}

\layoutstyle{6x9}

\usepackage{graphicx}
\usepackage{bm}       % bold math
\usepackage{mathptmx} % postscript fonts
\begin{document}

\long \def \blockcomment #1\endcomment{}
 
\title{The Early Days of Lattice Gauge Theory}

\author{Michael Creutz}
{address={
Physics Department, Brookhaven National Laboratory\\
Upton, NY 11973, USA
}
}

\thanks{This manuscript has been authored under
contract number DE-AC02-98CH10886 with the U.S.~Department of Energy.
Accordingly, the U.S. Government retains a non-exclusive, royalty-free
license to publish or reproduce the published form of this
contribution, or allow others to do so, for U.S.~Government purposes.}

\begin{abstract}
I discuss some of the historical circumstances that drove us to use
the lattice as a non-perturbative regulator.  This approach has had
immense success, convincingly demonstrating quark confinement and
obtaining crucial properties of the strong interactions from first
principles.  I wrap up with some challenges for the future.
\end{abstract}

\maketitle

\section{Introduction}

I am honored to have this opportunity to talk at this meeting in honor
of Nicholas Metropolis.  His historic work has played a crucial role
in many fields, and was absolutely crucial to the development of
lattice gauge theory.  In this talk I will reminisce a bit about the
early days, trying to explain why a technique from a rather different
field became such a crucial tool to the particle theory community.  I
will summarize some of the successes and mention a few unsolved
problems.

\section {Particle physics before the lattice}

I begin by summarizing the situation in particle physics in the late
60's, when I was a graduate student.  Quantum-electrodynamics had
already been immensely successful, but that theory was in some sense
``done.''  While hard calculations remained, and indeed still remain,
there was no major conceptual advance remaining.

These were the years when the ``eightfold way'' for describing
multiplets of particles had recently gained widespread acceptance.
The idea of ``quarks'' was around, but with considerable caution about
assigning them any physical reality; maybe they were nothing but a
useful mathematical construct.  A few insightful theorists were
working on the weak interactions, and the basic electroweak
unification was beginning to emerge.  The SLAC experiments were
observing substantial inelastic electron-proton scattering at large
angles, and this was quickly interpreted as evidence for substructure,
with the term ``parton'' coming into play.  While occasionally there
were speculations relating quarks and partons, people tended to be
rather cautious about pushing this too hard.

A crucial feature at the time was that the extension of quantum
electrodynamics to a meson-nucleon field theory was failing miserably.
The analog of the electromagnetic coupling had a value about 15, in
comparison with the 1/137 of QED.  This meant that higher order
corrections to perturbative processes were substantially larger than
the initial calculations.  There was no known small parameter in which
to expand.

In frustration over this situation, much of the particle theory
community set aside traditional quantum field theoretical methods and
explored the possibility that particle interactions might be
completely determined by fundamental postulates such as analyticity
and unitarity.  This ``S-matrix'' approach raised the deep question of
just ``what is elementary.''  A delta baryon might be regarded as a
combination of a proton and a pion, but it would be just as correct to
regard the proton as a bound state of a pion with a delta.  All
particles are bound together by exchanging themselves.  These ``dual''
views of the basic objects of the theory persist today in string
theory.

As we entered the 1970's, partons were increasingly identified with
quarks.  This shift was pushed by two dramatic theoretical
accomplishments.  First was the proof of renormalizability for
non-Abelian gauge theories \cite{renormalizability}, giving confidence
that these elegant mathematical structures \cite{ym} might have
something to do with reality.  Second was the discovery of asymptotic
freedom, the fact that interactions in non-Abelian theories become
weaker at short distances \cite{asymptoticfreedom}.  Indeed, this was
quickly connected with the point-like structures hinted at in the SLAC
experiments.  Out of these ideas evolved QCD, the theory of quark
confining dynamics.

The viability of this picture depended upon the concept of
``confinement.''  While there was strong evidence for quark
substructure, no free quarks were ever observed.  This was
particularly puzzling given the nearly free nature of their apparent
interactions inside the nucleon.  This returns us to the question of
``what is elementary?''  Are the fundamental objects the physical
particles we see in the laboratory or are they these postulated quarks
and gluons?

\begin{figure}
\centering
\includegraphics[width=.5\hsize]{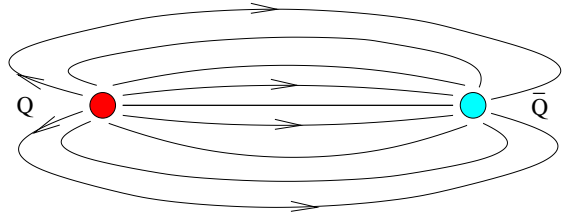}
\caption {A tube of gluonic flux connects quarks and anti-quarks.  The
strength of this string is 14 tons.}
\label{fluxtube}
\end{figure}

Struggling with this paradox led to the now standard flux-tube picture
of confinement.  The gluons are analogues of photons except that they
carry ``charge'' with respect to each other.  Massless charged
particles are rather singular objects, leading to a conjectured
instability that removes zero mass gluons from the spectrum, but does
not violate Gauss's law.  A Coulombic $1/r^2$ field is a solution of
the equations of a massless field, but, without massless particles,
such a spreading of the gluonic flux is not allowed.  The field lines
from a quark cannot end, nor can they spread in the inverse square law
manner.  Instead, as in Fig.~{\ref{fluxtube}}, the flux lines cluster
together, forming a tube emanating from the quark and ultimately
ending on an anti-quark.  This structure is a real physical object, and
grows in length as the quark and anti-quark are pulled apart.  The
resulting force is constant at long distance, and is measured via the
spectrum of high angular momentum states, organized into the famous
``Regge trajectories.''  In physical units, the flux tube pulls with a
strength of about 14 tons.

The reason a quark cannot be isolated is similar to the reason that a
piece of string cannot have just one end.  Of course one can't have a
piece of string with three ends either, but this is the reason for the
underlying $SU(3)$ group theory.  The confinement phenomenon cannot be
seen in perturbation theory; when the coupling is turned off, the
spectrum becomes free quarks and gluons, dramatically different than
the pions and protons of the interacting theory.

The mid 70's marked a particularly exciting time for particle physics,
with a series of dramatic events revolutionizing the field.  First was
the discovery of the $J/\psi$ particle \cite{jpsi}.  The
interpretation of this object and its partners as bound states of
heavy quarks provided the hydrogen atom of QCD.  The idea of quarks
became inescapable; field theory was reborn.  The $SU(3)$ non-Abelian
gauge theory of the strong interactions was combined with the
electroweak theory to become the durable ``standard model.''

This same period also witnessed several remarkable realizations on the
more theoretical front.  Non-linear effects in classical field
theories were shown to have deep consequences for their quantum
counterparts.  Classical ``lumps'' represented a new way to get
particles out of a quantum field theory \cite{lumps}.  Much of the
progress here was in two dimensions, where techniques such as
``bosonization'' showed equivalences between theories of drastically
different appearance.  A boson in one approach might appear as a bound
state of fermions in another, but in terms of the respective
Lagrangian approaches, they were equally fundamental.  Again, we were
faced with the question ``what is elementary?''  Of course modern
string theory is discovering multitudes of ``dualities'' that continue
to raise this same question.

These discoveries had deep implications: field theory can have much
more structure than seen from the traditional analysis of Feynman
diagrams.  But this in turn had crucial consequences for practical
calculations.  Field theory is notorious for divergences requiring
regularization.  The bare mass and charge are infinite quantities.
They are not the physical observables, which must be defined in terms
of physical processes.  To calculate, a ``regulator'' is required to
tame the divergences, and when physical quantities are related to each
other, any regulator dependence should drop out.

The need for controlling infinities had, of course, been known since
the early days of QED.  But all regulators in common use were based on
Feynman diagrams; the theorist would calculate diagrams until one
diverged, and that diagram was then cut off.  Numerous schemes were
devised for this purpose, ranging from the Pauli-Villars approach to
forest formulae to dimensional regularization.  But with the
increasing realization that non-perturbative phenomena were crucial,
it was becoming clear that we needed a ``non-perturbative'' regulator,
independent of diagrams.

\section{The lattice}

The necessary tool appeared with Wilson's lattice theory.  He
originally presented this as an example of a model exhibiting
confinement.  The strong coupling expansion has a non-zero radius of
convergence, allowing a rigorous demonstration of confinement, albeit
in an unphysical limit.  The resulting spectrum has exactly the
desired properties; only gauge singlet bound states of quarks and
gluons can propagate.

This was not the first time that the basic structure of lattice gauge
theory had been written down.  A few years earlier, Wegner
\cite{wegner} presented a $Z_2$ lattice gauge model as an example of a
system possessing a phase transition but not exhibiting any local
order parameter.  In his thesis, Jan Smit \cite{smit} described using
a lattice regulator to formulate gauge theories outside of
perturbation theory.  The time was clearly ripe for the development of
such a regulator.  Very quickly after Wilson's suggestion, Balian,
Drouffe, and Itzykson \cite{bdi} explored an amazingly wide variety of
aspects of these models.

To reiterate, the primary role of the lattice is to provide a
non-perturbative cutoff.  Space is not really meant to be a crystal,
the lattice is a mathematical trick.  It provides a minimum wavelength
through the lattice spacing $a$, {\it i.e.} a maximum momentum of
$\pi/a$.  Path summations become well defined ordinary integrals.  By
avoiding the convergence difficulties of perturbation theory, the
lattice provides a route to the rigorous definition of quantum field
theory.

The approach, however, had a marvelous side effect.  By discreetly
making the system discrete, it becomes sufficiently well defined to be
placed on a computer.  This was fairly straightforward, and came at
the same time that computers were growing rapidly in power.  Indeed,
numerical simulations and computer capabilities have continued to grow
together, making these efforts the mainstay of lattice gauge theory.

Now I wish to reiterate one of the most remarkable aspects of the
theory of quarks and gluons, the paucity of adjustable parameters.  To
begin with, the lattice spacing itself is not an observable.  We are
using the lattice to define the theory, and thus for physics we are
interested in the continuum limit $a\rightarrow 0$.  Then there is the
coupling constant, which is also not a physical parameter due to the
phenomenon of asymptotic freedom.  The lattice works directly with a
bare coupling, and in the continuum limit this should vanish
$$
e_0^2 \rightarrow 0
$$ 
In the process, the coupling is replaced by an overall scale.  Coleman
and Weinberg \cite{colemanweinberg} gave this phenomenon the marvelous
name ``dimensional transmutation.''  Of course an overall scale is not
really something we should expect to calculate from first principles.
Its value would depend on the units chosen, be they furlongs or
light-fortnights.
 
Next consider the quark masses.  Indeed, measured in units of the
asymptotic freedom scale, these are the only free parameters in the
strong interactions.  Their origin remains one of the outstanding
mysteries of particle physics.  The massless limit gives a rather
remarkable theory, one with no undetermined dimensionless parameters.
This limit is not terribly far from reality; chiral symmetry breaking
should give massless pions, and experimentally the pion is
considerably lighter than the next non-strange hadron, the rho.  A
theory of two massless quarks is a fair approximation to the strong
interactions at intermediate energies.  In this limit all
dimensionless ratios should be calculable from first principles,
including quantities such as the rho to nucleon mass ratio.

The strong coupling at any physical scale is not an input parameter,
but should be determined.  Such a calculation has gotten lattice gauge
theory into the famous particle data group tables \cite{pdg}.  With
appropriate definition the current lattice result is
$$
\alpha_s(M_Z)=0.115\pm 0.003
$$
where the input is details of the charmonium spectrum.

\section{Numerical simulation}

While other techniques exist, large scale numerical simulations
currently dominate lattice gauge theory.  They are based on attempts
to evaluate the path integral
$$
Z=\int dU e^{-\beta S}
$$
with $\beta$ proportional to the inverse bare coupling squared.  A
direct evaluation of such an integral has pitfalls.  At first sight,
the basic size of the calculation is overwhelming.  Considering a
$10^4$ lattice, small by today standards, there are 40,000 links.  For
each is an $SU(3)$ matrix, parametrized by 8 numbers.  Thus we have a
$10^4\times 4 \times 8 = 320,000$ dimensional integral.  One might try
to replace this with a discrete sum over values of the integrand.  If
we make the extreme approximation of using only two points per
dimension, this gives a sum with
$$
2^{320,000}=3.8\times 10^{96,329}
$$
terms!  Of course, computers are getting pretty fast, but one should
remember that the age of universe is only $\sim 10^{27}$ nanoseconds.

These huge numbers suggest a statistical treatment.  Indeed, the above
integral is formally just a partition function.  Consider a more
familiar statistical system, such as a glass of beer.  There are a
huge number of ways of arranging the atoms of carbon, hydrogen,
oxygen, etc.~that still leaves us with a glass of beer.  We don't need
to know all those arrangements, we only need a dozen or so ``typical''
glasses to know all the important properties.

This is the basis of the Monte Carlo approach.  The analogy with a
partition function and the role of ${1\over \beta}$ as a temperature
enables the use of standard techniques to obtain ``typical''
equilibrium configurations, where the probability of any given
configuration is given by the Boltzmann weight
$$
P(C)\sim e^{-\beta S(C)}
$$
For this we use a Markov process, making changes in the current
configuration
$$
C\rightarrow C^\prime \rightarrow \ldots
$$
biased by the desired weight.

The idea is easily demonstrated with the example of $Z_2$ lattice
gauge theory \cite{creutzjacobsrebbi}.  For this toy model the links
are allowed to take only two values, either plus or minus unity.
One sets up a loop over the lattice variables.  When looking at a
particular link, calculate the probability for it to have value $1$
$$
P(1)={e^{-\beta S(1)}\over e^{-\beta S(1)}+e^{-\beta S(-1)}}
$$ 
Then pull out a roulette wheel and select either 1 or $-1$ biased by
this weight.  Lattice gauge Monte-Carlo programs are by nature quite
simple.  They are basically a set of nested loops surrounding a random
change of the fundamental variables.

\begin{figure}
\centering
\includegraphics[width=.85\hsize]{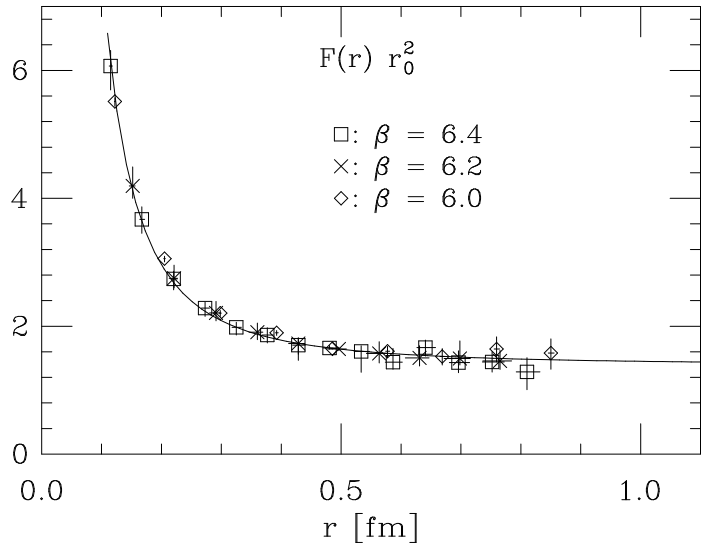}
\caption{The force between two quarks does not fall to zero as the
distance increases.  This is the confinement phenomenon. (From
Ref.~\cite{michael}).}
\label{force}
\end{figure}

The results of these simulations have been fantastic, giving first
principles calculations of interacting quantum field theory.  I will
just mention two examples.  The early result that bolstered the
lattice into mainstream particle physics was the convincing
demonstration of the confinement phenomenon.  The force between two
quark sources indeed remains constant at large distances.  A summary
of this result is shown in Fig.~\ref{force}, taken from
Ref.~\cite{michael}.

% C. Michael, hep-lat/9509090 

\begin{figure}
\centering
\includegraphics[width=.7\hsize,clip]{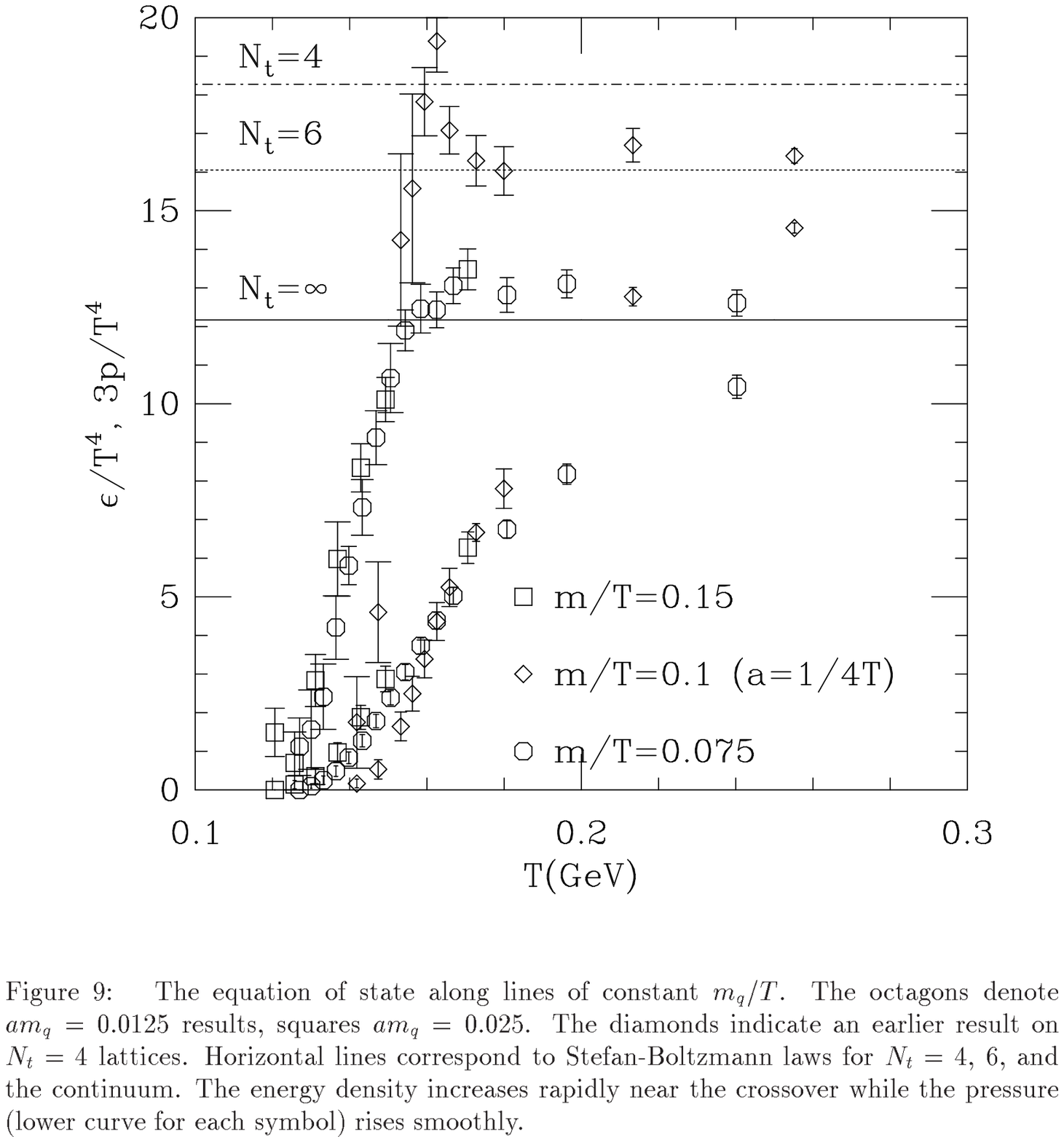}
\caption{The energy and pressure of the {\ae}ther show a dramatic
structure at a temperature of about 170--190 MeV.  The lattice is a unique
theoretical tool for the study of this transition to a quark-gluon
plasma (From Ref.~\cite{eos}).}
\label{finitetemp}
\end{figure}

Another accomplishment for which the lattice excels over all other
methods has been the study the deconfinement of quarks and gluons into
a plasma at a temperature of about 170--190 Mev\cite{plasma}.  Indeed,
the lattice is a unique quantitative tool capable of making precise
predictions for this temperature.  The method is based on the fact
that the Euclidean path integral in a finite temporal box directly
gives the physical finite temperature partition function, where the
size of the box is proportional to the inverse temperature.  This
transition represents a loss of confining flux tubes in a background
plasma.  Fig.~\ref{finitetemp} shows one calculation of this
transition \cite{eos}.

\section{Quarks}

While the gauge sector of the lattice theory is in good shape, from
the earliest days fermionic fields have caused annoying difficulties.
Actually there are several apparently unrelated fermion problems.  The
first is an algorithmic one.  The quark operators are not ordinary
numbers, but anti-commuting operators in a Grassmann space.  As such
the exponentiated action itself is an operator.  This makes comparison
with random numbers problematic.

Over the years various clever tricks for dealing with this problem
have been developed; numerous large scale Monte Carlo simulations do
involving dynamical fermions.  The algorithms used are all essentially
based on an initial analytic integration of the quarks to give a
determinant.  This, however, is the determinant of a rather large
matrix, the size being the number of lattice sites times the number of
fermion field components, with the latter including spinor, flavor,
and color factors.  In my opinion, the algorithms working directly
with these large matrices remain quite awkward.  I often wonder if
there is some more direct way to treat fermions without the initial
analytic integration.

The algorithmic problem becomes considerably more serious when a
chemical potential generating a background baryon density is present.
In this case the required determinant is not positive; it cannot be
incorporated as a weight in a Monte Carlo procedure.  This is
particularly frustrating in the light of striking predictions of
super-conducting phases at large chemical potential
\cite{superconduct}.  This is perhaps the most serious unsolved
problem in lattice gauge theory.

The other fermion problems concern chiral issues.  There are a variety
of reasons that such symmetries are important in physics.  First is
the light nature of the pion, which is traditionally related to the
spontaneous breaking of a chiral symmetry expected to become exact as
the quark masses go to zero.  Second, the standard model itself is
chiral, with the weak bosons coupling to chiral currents.  Third, the
idea of chiral symmetry is frequently used in the development of
unified models as a tool to prevent the generation of large masses and
thus avoid fine tuning.

Despite its importance, chiral symmetry and the lattice have never fit
particularly well together.  I regard this as evidence that the
lattice is trying to tell us something deep.  Indeed, the lattice
fully regulates the theory, and thus all the famous anomalies must be
incorporated explicitly.  It is well known that the standard model is
anomalous if either the quarks or leptons are left out, and this
feature must appear in any valid formulation.

These issues are currently a topic with lots of activity
\cite{myreview}.  Several schemes for making chiral symmetry more
manifest have been developed, with my current favorite being the
domain-wall formulation, where our four dimensional world is an
interface in an underlying five dimensional theory.

\section{The Lattice SciDAC Project}

Lattice gauge theory has grown into a powerful tool.  Indeed, it is
becoming essential to the interpretation of experiments at all the
high energy and nuclear physics laboratories.  But in many cases the
theoretical errors dominate, and we need improved computing resources
for further progress.  Realizing the need to work together on this,
the US lattice gauge community has put together a collaborative effort
towards the goal of providing terascale computing resources.
Currently 66 US lattice theorists are signed on and have set up a 9
member executive committee, chaired by R. Sugar of UC Santa Barbara
and including myself as a member.  We are proposing a two pronged
approach, with a next generation special purpose machine to be based
at Brookhaven Lab, and two large scale commodity clusters to be based
at Fermilab and Jefferson Lab.  The goal is to have in a few years
three 10 teraflops scale resources available to the community.

The machine to go at Brookhaven is called the QCDOC for ``QCD on a
chip.''  A single node is designed into a single application specific
integrated circuit, designed in collaboration between Columbia
University and IBM.  These will be integrated into a six dimensional
mesh.  The RIKEN/BNL Research Center and the UKQCD collaboration have
each ordered 5 teraflops sustained versions of this machine.  The hope
is to have in addition a DOE sponsored 10 teraflops sustained QCDOC
for the US community by the end of 2004.

In conjunction with this project is a software effort to make these
machines easily accessible to the community.  We want the same software
at the top level to run with minimal modifications on all machines,
including both the clusters and the QCDOC.  More information on this
project can be found at www.lqcd.org. 

\section{Concluding remarks}

I conclude by mentioning two problems that particularly interest me.
These are all directly connected with the problems of quarks.  The
first is the chiral symmetry problem, alluded to above.  Here the
recent developments have put parity conserving theories, such as the
strong interactions, into quite good shape.  The various schemes,
including domain-wall fermions, the overlap formula, and variants on
the Ginsparg-Wilson relation, all quite elegantly give the desired
chiral properties.  Chiral gauge theories themselves, such as the weak
interactions, are not yet completely resolved, but the above
techniques appear to be tantalizingly close to a well defined lattice
regularization.  It is still unclear whether the lattice
regularization can simultaneously be fully finite, gauge invariant,
and local.  The problems encountered are closely related to similar
issues with super-symmetry, another area that does not naturally fit on
the lattice.  This also ties in with the explosive activity in string
theory and a possible regularization of gravity.

The other area in particular need of advancement lies in dynamical
fermion methods.  As I said earlier, I regard all existing algorithms
as frustratingly awkward.  This, plus the fact that the sign problem
with a background density remains unsolved, suggests that new ideas
are needed.  It has long bothered me that we treat fermions and bosons
so differently in numerical simulations.  Indeed, why do we have to
treat them separately?

\end{document}